\documentclass[cup6b,epsf]{cupbook}

\usepackage{epsf}
\def\be{\begin{equation}}
\def\ee{\end{equation}}
\def\simgreat{\mathbin{\lower 3pt\hbox
   {$\rlap{\raise 5pt\hbox{$\char'076$}}\mathchar"7218$}}}
\def\simless{\mathrel{\raise.3ex\hbox{$<$}\mkern-14mu
             \lower0.6ex\hbox{$\sim$}}}
\def\go{\mathrel{\raise.3ex\hbox{$>$}\mkern-14mu
             \lower0.6ex\hbox{$\sim$}}}
\def\lo{\mathrel{\raise.3ex\hbox{$<$}\mkern-14mu
             \lower0.6ex\hbox{$\sim$}}}
\def\kms{{\rm km\,s}^{-1}}

\begin{document}

\author{Dong Lai\\
Center for Radiophysics and Space Research, Department of Astronomy\\
Cornell University, Ithaca, NY 14853}

\chapter{Neutron Star Kicks and Supernova Asymmetry}

{\it Observations over the last decade have shown that
neutron stars receive a large kick velocity (of order a few hundred to a
thousand km\,s$^{-1}$) at birth. The physical origin of the kicks 
and the related supernova asymmetry is one of the central
unsolved mysteries of supernova research. We review the physics of 
different kick mechanisms, including hydrodynamically driven, 
neutrino -- magnetic field driven, and electromagnetically driven kicks. 
The viabilities of the different kick mechanisms are directly related to 
the other key parameters characterizing nascent neutron stars, such as
the initial magnetic field and the initial spin. Recent
observational constraints on kick mechanisms are also discussed.}

\section{Evidence for Neutron Star Kicks and Supernova Asymmetry}

It has long been recognized that neutron stars (NSs) have space velocities 
much greater than their progenitors'.
A natural explanation for such high velocities is that supernova (SN)
explosions are asymmetric, and provide kicks to the nascent NSs.
In recent years evidence for NS kicks and NS asymmetry 
has become much stronger. The observations that support 
(or even require) NS kicks fall into three categories:

{\it (1) Large NS Velocities ($\gg$ the progenitors' velocities 
$\sim 30$~km\,s$^{-1}$):} 

$\bullet$ The study of pulsar proper motion give a mean birth 
velocity $200-500$~km\,s$^{-1}$ 
(Lorimer et al.~1997; Hansen \& Phinney 1997; 
Cordes \& Chernoff 1998;
Arzoumanian et al~2002), with possibly a significant population having
$V\go 1000$~km\,s$^{-1}$. 
While velocity of $\sim 100$~km\,s$^{-1}$ may in principle
come from binary breakup in a supernova (without kick), higher velocities 
would require exceedingly tight presupernova binary. Statistical analysis
seems to favor a bimodal pulsar velocity distribution, with peaks
around $100~\kms$ and $500~\kms$ (see Arzoumanian et al.~2002). 

$\bullet$ Observations of bow shock from the Guitar nebula
pulsar (B2224+65) implies $V \simgreat 1000$~km~s$^{-1}$
(Cordes et al.~1993; Chatterjee \& Cordes 2002).

$\bullet$ The studies of NS -- SNR associations have, 
in some cases, implied large NS velocities, up to $\sim 10^3$~km~s$^{-1}$
(e.g., NS in Cas A SNR has $V>330$~km~s$^{-1}$; Thorstensen et al.~2001).

{\it (2) Characteristics of NS Binaries:}
While large space velocities can in principle be accounted for 
by binary break-up 
(see Iben \& Tutukov 1996), many observed characteristics of NS binaries 
demonstrate that binary break-up can not be solely responsible for 
pulsar velocities, and that kicks are required. 
Examples include:

$\bullet$ The spin-orbit misalignment in PSR J0045-7319/B-star binary,
as manifested by the orbital plane precession 
(Kaspi et al.~1996; Lai et al.~1995) and fast orbital decay (which indicates
retrograde rotation of the B star with respect to the orbit; Lai 1996;
Kumar \& Quataert 1997) require that the NS received a kick at birth
(see Lai 1996b). Similar precession of orbital plane has been observed in PSR
J1740-3052 system (Stairs et al.~2003).

$\bullet$ The detection of geodetic precession in binary pulsar PSR 1913+16 
implies that the pulsar's spin is misaligned with the orbital angular momentum;
this can result from the aligned pulsar-He star progenitor only if
the explosion of the He star gave a kick to the NS that misalign 
the orbit (Kramer 1998; Wex et al.~1999). 

$\bullet$ The system radial velocity ($430\,\kms$) of X-ray
binary Circinus X-1 requires $V_{\rm kick}\simgreat 500$\,km\,s$^{-1}$ 
(Tauris et al.~1999). Also, PSR J1141-6545 has $V_{\rm sys}\simeq 125~\kms$.

$\bullet$ High eccentricities of Be/X-ray binaries cannot be
explained without kicks 
(van den Heuvel \& van Paradijs 1997; 
but see Pfahl et al.~2002).

$\bullet$ Evolutionary studies of NS binary population 
(in particular the double NS systems)
imply the existence of pulsar kicks (e.g., 
Fryer et al.~1998). 

{\it (3) Observations of SNe and SNRs:}
There are many direct observations, detailed in other contributions to this 
proceedings, of nearby supernovae 
(e.g., spectropolarimetry, Wang et al.~2003;
X-ray and gamma-ray observations and emission line profiles of SN1987A)
and supernova remnants 
which support the notion that supernova explosions are not spherically
symmetric.

\section{The Problem of Core-Collapse Supernovae and NS Kicks}

The current paradigm for core-collapse supernovae leading to NS formation
is that these supernovae are neutrino-driven (see, e.g.,
Burrows \& Thompson 2002; Janka et al.~2002 for a recent review): 
As the central core of a massive star collapses to nuclear density, it rebounds
and sends off a shock wave, leaving behind a proto-NS. The shock
stalls at several 100's km because of neutrino loss and
nuclear dissociation in the shock. A fraction of 
the neutrinos emitted from the proto-NS get absorbed
by nucleons behind the shock, thus reviving the shock, 
leading to an explosion on the timescale several 100's ms ---  
This is the so-called ``delayed mechanism''. 
It has been argued that neutrino-driven convection in the proto-NS
and that in the shocked mantle
are central to the explosion mechanism (e.g., Mezzacappa et al.~1998), 
although current 2D simulations with the state-of-the-art neutrino interaction
and transport have not produced a successful explosion model
(Buras et al.~2003; see Fryer \& Warren 2003 for simulations in 3D
that use more approximate neutrino physics/transport). 
What is even more uncertain is
the role of rotation and magnetic field on the explosion (see 
Rampp, M\"uller \& Ruffert 1998; Fryer \& Heger 2000; Ott et al.~2004
for simulations of 
collapse/explosion with rotation, and Thompson \& Norman 2001,
Wheeler et al.~2002 and 
Akiyama et al.~2003 
and references therein for discussions of magnetic effects).

It is clear that 
our understanding of the physical mechanisms of core-collapse supernovae
remains rather incomplete. The prevalence of neutron star kicks poses a
significant mystery, and indicates that large-scale, global deviation from
spherical symmetry is an important ingredient in our understanding of
core-collapse supernovae.
In the following sections, we review different classes of 
physical mechanisms for generating 
NS kicks, and then discuss possible 
observational constraints and astrophysical implications.

\section{Kick Mechanisms}

\subsection{Hydrodynamically Driven Kicks} 

{\it (1) Can Convections lead to NS Kicks?}
The collapsed stellar core and its surrounding mantle are susceptible
to a variety of hydrodynamical (convective) instabilities 
(e.g., Herant et al.~1994; Burrows et al.~1995; Janka \& M\"uller 1996;
Mezzacappa et al.~1998). It is natural to expect that 
the asymmetries in the density, temperature and velocity distributions 
associated with the instabilities can lead to asymmetric matter ejection 
and/or asymmetric neutrino emission. Most numerical simulations in the 1990s 
indicate that the local, post-collapse instabilities are not adequate
to account for kick velocities $\simgreat 100$~km~s$^{-1}$.
Recently, Scheck et al.~(2003) reported computer experiments in which they
adjust the neutrino luminosity from the neutrinosphere (the inner boundary 
of the simulation domain) to obtain successful explosions. They found that for
long-duration (more than a second) explosions, neutrino-driven convection
behind the expanding shock can lead to global ($l=1,2$) asymmetries,
accelerating the remnant NS to a range of velocities up to several 
hundreds of $\kms$ (cf.~Thompson 2000). 
This result is encouraging. But note that,
like most other SN simulations, the Scheck et al. simulations 
were done in 2D, the proton-NS was fixed
on the grids (with the kick calculated by adding up the momentum
flux across the inner boundary), and the explosions were obtained in an ad hoc 
manner. It is also not clear whether kick velocities of 500-1000~$\kms$ 
can be easily obtained. 

\begin{figure}
\begin{center}
\leavevmode\epsfxsize=11cm \epsfbox{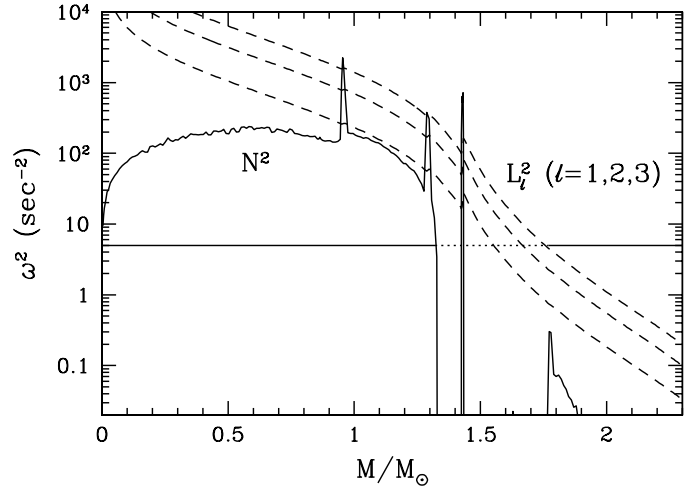}
\end{center}
\vskip -3.4cm
\caption{
Propagation diagram computed for a $15M_\odot$
presupernova model of Weaver and Woosley (1993). 
The solid curve shows $N^2$, where $N$ is the Brunt-V\"ais\"al\"a
frequency; the dashed curves show $L_l^2$, where $L_l$ is the 
acoustic cutoff frequency, with $l=1,~2,~3$.
The spikes in $N^2$ result from discontinuities in entropy and
composition. The iron core boundary is located at $1.3M_\odot$,
the mass-cut at $1.42M_\odot$.
Convective regions correspond to $N=0$. Gravity modes
(with mode frequency $\omega$) propagate in regions
where $\omega<N$ and $\omega<L_l$, while pressure modes
propagate in regions where $\omega>N$ and $\omega>L_l$.
Note that a g-mode trapped in the core can lose energy
by penetrating the evanescent zones and turning into an
outgoing acoustic wave (see the horizontal line). 
Also note that g-modes with higher $n$ (the
radial order) and $l$ (the angular degree) are better trapped in the
core than those with lower $n$ and $l$.
}
\label{fig1}
\end{figure}

{\it (2)Asymmetries in Pre-Supernova Cores:}
It has been recognized that one way to produce large kicks is to have 
global asymmetric perturbations prior to core collapse
(Goldreich et al.~1996; Burrows \& Hayes 1996).
One possible origin for the pre-SN asymmetry is the
overstable oscillations in the pre-SN core (Goldreich et al.~1996).
The idea is the following. A few hours prior to core collapse,
the central region of the progenitor star consists of a Fe
core surrounded by Si-O burning shells and other layers of envelope.
This configuration is overstable to nonspherical oscillation modes.
It is simplest to see this by considering a $l=1$ mode: If we perturb
the core to the right, the right-hand-side of the shell will be compressed,
resulting in an increase in
temperature; since the shell nuclear burning rate depends sensitively
on temperature (power-law index $\sim 47$ for Si burning and $\sim 33$ for O
burning), the nuclear burning is greatly enhanced; this generates a large
local pressure, pushing the core back to the left.
The result is an oscillating g-mode with increasing amplitude.
There are also damping mechanisms for these modes, the most important
one being leakage of mode energy:
The local (WKB) dispersion relation for nonradial waves is 
\be
k_r^2=(\omega^2c_s^2)^{-1}(\omega^2-L_l^2)(\omega^2-N^2),
\ee
where $k_r$ is the radial wavenumber, $L_l=\sqrt{l(l+1)}c_s/r$ ($c_s$ 
is the sound speed) and $N$ are the acoustic cut-off (Lamb) frequency and the 
Brunt-V\"ais\"al\"a frequency, respectively.
Since acoustic waves whose frequencies lie above the acoustic cutoff can
propagate through convective regions, each core g-mode will couple to an
outgoing acoustic wave, which drains energy from the core g-modes
(see Fig.~\ref{fig1}).
In another word, the g-mode is not exactly trapped in the core.
Our calculations (based on the $15M_\odot$ and $25M_\odot$
presupernova models of Weaver \& Woosley) indicate that
a large number of g-modes are overstable, although for low-order modes
(small $l$ and $n$) the results depend sensitively on the detailed
structure and burning rates of the presupernova models (see Lai 2001).
The typical mode periods are $\simgreat 1$~s, the growth time
$\sim 10-50$~s, and the lifetime of the Si shell burning is $\sim$ hours.
Thus there could be a lot of e-foldings for the nonspherical
g-modes to grow. Our preliminary calculations based on the recent models of
A. Heger and S. Woosley (Heger et al.~2001)
give similar results (work in progress).
Our tentative conclusion is that overstable g-modes may potentially grow
to large amplitudes prior to core implosion, although several issues
remain to be understood better.
For example, the O-Si burning shell is highly convective,
with convective speed reaching $1/4$ of the sound speed, and
hydrodynamical simulation may be needed to properly modeled such convection
zones (see 
Asida \& Arnett 2000).

So now we have a plausible way of generating initial asymmetric perturbations
before core collapse. During the collapse, the asymmetries
are amplified by a factor of 5-10 (Lai \& Goldreich 2000; see also Lai 2000).
How do we get the kick? The numerical simulations by
Burrows \& Hayes (1996) illustrate the effect.
Suppose the right-hand-side
of the collapsing core is denser than the left-hand side. As the shock wave
comes out after bounce, it will see different densities in different
directions, and it will move preferentially on the direction
where the density is lower. So we have an asymmetric shock
propagation and mass ejection, a ``mass rocket''
(cf. Fryer 2003).

\subsection{Neutrino -- Magnetic Field Driven Kicks}

The second class of kick mechanisms rely on asymmetric neutrino emission
induced by strong magnetic fields. 
Since $99\%$ of the NS binding energy (a few times $10^{53}$~erg) is
released in neutrinos, tapping the neutrino energy would appear to be an
efficient means to kick the newly-formed NS.
The fractional asymmetry $\alpha$ in the radiated neutrino energy required to
generate a kick velocity $V_{\rm kick}$ is $\alpha=MV_{\rm kick}c/E_{\rm tot}$
($=0.028$ for $V_{\rm kick}=1000$~km~s$^{-1}$, NS mass
$M=1.4\,M_\odot$ and total neutrino energy radiated $E_{\rm tot}
=3\times 10^{53}$~erg). There are several possible effects:

{\it (1) Parity Violation:} Because weak interaction is parity
violating, the neutrino opacities and emissivities in a magnetized nuclear
medium depend asymmetrically on the directions of neutrino momenta with respect
to the magnetic field, and  this can give rise to asymmetric neutrino emission
from the proto-NS. Calculations indicate that to generate interesting kicks
with this effect requires the proto-NS to have a large-scale, ordered
magnetic field of at least a few $\times 10^{15}$~G (see Arras \& Lai 1999a,b
and references therein).

{\it (2) Asymmetric Field Topology:}
Another effect relies on the asymmetric magnetic field
distribution in proto-NSs: Since the cross section for
$\nu_e$ ($\bar\nu_e$) absorption on neutrons (protons) depends on the local
magnetic field strength, the local neutrino fluxes emerged from
different regions of the stellar surface are different. Calculations
indicate that to generate a kick velocity of $\sim 300$~km~s$^{-1}$ using this
effect alone would require that the difference
in the field strengths at the two opposite poles of the star
be at least $10^{16}$~G (see Lai \& Qian 1998). Note that
only the magnitude of the field matters here.

{\it (3) Dynamical Effect of Magnetic Fields:}
A superstrong magnetic field may also play a dynamical role in the
proto-NS. For example, it has been suggested that a locally strong
magnetic field can induce ``dark spots'' (where the neutrino flux is
lower than average) on the stellar surface by suppressing
neutrino-driven convection (Duncan \& Thompson 1992). While it is
difficult to quantify the kick velocity resulting from an asymmetric
distribution of dark spots, order-of-magnitude estimate indicates that a local
magnetic field of at least $10^{15}$~G is needed for this effect to be
of importance.


{\it (4) Exotic Neutrino Physics:}
There have also been several ideas of pulsar kicks
which rely on nonstandard neutrino physics. For example, it was
suggested (Kusenko \& Segre 1996, 1998) 
that asymmetric $\nu_\tau$ emission could result from the MSW flavor
transformation between $\nu_\tau$ and $\nu_e$ inside a magnetized proto-NS
because a magnetic field changes the resonance condition for MSW effect.
This mechanism requires neutrino mass of order
$100$~eV. A similar idea (Akhmedov et al.~1997; Grasso et al.~1998)
relies on both the neutrino mass and the neutrino magnetic 
moment to facilitate the flavor transformation (resonant neutrino
spin-flavor precession).
Fuller et al.~(2003) discussed the effect of sterile neutrinos.
Analysis of neutrino transport (Janka \& Raffelt 1998) indicates
that even with favorable neutrino parameters, strong magnetic fields
$B\gg 10^{15}$~G are required to obtain a $100$~km~s$^{-1}$ kick.

\subsection{Electromagnetically Drievn Kicks}

Harrison \& Tademaru (1975) showed that electromagnetic (EM) radiation
from an off-centered rotating magnetic dipole imparts
a kick to the pulsar along its spin axis. The kick is attained
on the initial spindown timescale of the pulsar (i.e.,
this really is a gradual acceleration), and
comes at the expense of the spin kinetic energy. A reexamination
of this effect (Lai et al.~2001) showed that the force on the
pulsar due to asymmetric EM radiation is larger than the original
Harrison \& Tademaru expression by a factor of four. Thus, the 
maximum possible velocity is $V_{\rm kick}^{(\rm max)}\simeq
1400\,\left({1\,{\rm kHz}/P_i}\right)^2~\kms$. 
Nevertheless, to generate interesting kicks using this mechanism
requires the initial spin period $P_i$ 
of the NS to be less than 1-2~ms. Gravitational radiation may also affect the
net velocity boost.

\subsection{Other Possibilities}

(1) If rotation and magnetic fields play a dominate role in 
the explosion, bipolar jets may be produced. A slight asymmetry between
the two jets will naturally lead to large kick (e.g., Khokhlov et al.~1999;
Akiyama et al.~2003). While difficult to calculate, this is a serious
possibility given the increasing observational evidence for
bipolar explosions in many SNe (see other contributions to this proceedings).

(2) Colpi and Wasserman (2003) considered the formation of double
proton-NS binary in a rapidly rotating core collapse; the lighter NS
explodes after reaching its minimum mass limit (via mass transfer),
giving the remaining NS a large kick ($\sim 10^3\,\kms$). A
related suggestion relies on the coalescence of proto-NS binary
as providing the kick (Davies et al.~2002). The biggest uncertainty
for such scenarios is that it is not clear core fragmentation can
take place in the collapse (and numerical simulations seem to say no;
see Fryer \& Warren 2003).

\section{Astrophysical Constraints on Kick Mechanisms}

The review in previous section clearly shows that 
NS kick is not only a matter of curiosity, it is intimately connected to 
the other fundamental parameters of young NSs (initial spin and magnetic 
field). For example, the neutrino-magnetic field driven mechanisms 
are of relevance only for $B\simgreat 10^{15}$~G. While recent observations
have lent strong support that some neutron stars (``magnetars'')
are born with such a superstrong magnetic field,
it is not clear (perhaps unlikely) that ordinary radio pulsars 
(for which large velocities have been measured) had initial magnetic fields of
such magnitude.

One of the reasons that it has been difficult to pin down the kick
mechanisms is the lack of correlation between NS velocity and the other
properties of NSs. The situation has changed with the recent X-ray
observations of the compact X-ray nebulae of
the Crab and Vela pulsars, which have a two sided asymmetric jet
at a position angle coinciding with the position angle of the pulsar's
proper motion (Pavlov et al.~2000; Helfand et al.~2001).
The symmetric morphology of the nebula
with respect to the jet direction strongly suggests that the jet is
along the pulsar's spin axis. Analysis of the polarization angle
of Vela's radio emission corroborates this interpretation (Lai et al.~2001).
Thus, while statistical analysis
of pulsar population neither support nor rule out any
spin-kick correlation, at least for the Vela and Crab pulsars
(and perhaps for several other pulsars; see Ng \& Romani 2003),
the proper motion and the spin axis appear to be aligned.

The apparent alignment between the spin axis and proper motion
raises an interesting question: Under what
conditions is the spin-kick alignment expected for different kick
mechanisms? Let us look at the three classes of mechanisms discussed
before (Lai et al.~2001): (1) For the electromagnetically driven kicks,
the spin-kick slignment is naturally produced. (Again, note that
$P_i\sim 1-2$~ms is required to generate sufficiently large $V_{\rm kick}$).
(2) For the neutrino--magnetic field driven kicks: The kick is imparted to
the NS near the neutrinosphere (at 10's of km) on the neutrino diffusion time,
$\tau_{\rm kick}\sim 10$~seconds. As long as the initial spin period
$P_i$ is much less than a few seconds, spin-kick alignment is naturally
expected. (3) For the hydrodynamically driven kicks: because the kick is
imparted at a large radius ($\go 100$~km), to get effective rotational
averaging, we require that the rotation period at $\sim 100$~km to be shorter
than the kick timescale ($\sim 100$~ms). This translates to $P_{\rm NS}\lo
1$~ms, which means that rotation must be dynamically important
\footnote{It has been suggested that even with zero initial angular momentum,
aligned spin-kick may be possible if the kick is the result of many 
off-centered small thrusts which are appropriately oriented (Spruit 
\& Phinney 1998). However, in all the mechanisms studied so far,
the kick is the result of a single or a few thrusts, and one would expect
the final spin to be roughly perpendicular the kick.}.
On the otherhand, if rotation indeed plays a dynamically important role,
the basic collapse and explosion may be qualitatively different (e.g.,
core bounce may occur at subnuclear density, the explosion is
weaker and takes the form of two-sided jets (e.g. Khokhlov et al.~1999;
Fryer \& Heger 2000).
The possibility of a kick in such 
systems has not been studied, but it is conceivable 
that an asymmetric dipolar perturbation may be coupled
to rotation, thus producing spin-kick alignment.
Current observations, however, seem to suggest that most pulsars
are born rotating slowly ($P_i>10$~ms), 
with rotation palying a negligible role in the dynamics
(see also Heger et al.~2003 for preSN evolution of rotating stars).

Currently we do not know whether spin-kick alignment is
a generic feature of all pulsars; if it is, then it can provide
powerful constraint on the kick mechanisms and the SN explosion
mechanisms in general.

Finally, it is worth noting that 
recent observations showed that black hole (BH) formation can
be accompanied by SN explosion: The companion of
the BH X-ray binary GRO J1655-40 (Nova Sco) and that of SAX J1819.3-2525
(V4641 SGR) have high abundance of $\alpha$-elements (Israelian et al.~1999;
Orosz et al.~2001), 
which can only be produced in a SN explosion
(see Podsiadlowski et al.~2002). 
Apparently, the BH forms in an indirect process where a shock wave successfully
makes an explosion and a NS forms temporarily followed by fall-back,
or loss of angular momentum and thermal energy in the proto-NS
which then collapses to a BH. This indirect process may explain the
the relatively large space velocity of GRO J1655-40.

\bigskip\noindent
{\it Acknowledgements}: Support for this work is provided in part by 
NASA NAG 5-12034 and NSF AST 0307252. I thank my collaborators P. Arras,
D. Chernoff, J. Cordes, P. Goldreich, A. Heger, 
Y.-Z. Qian and A. Shirakawa for their important contributions. 
I also thank the conference organizers for a stimulating 
meeting and travel support.

\end{document}